\documentclass[12pt]{article}
\usepackage{geometry}
\usepackage{amsmath}
\usepackage{amssymb}
\usepackage{amsmath}
\usepackage{graphicx}
\numberwithin{equation}{section}

\def\appendix#1{
 \addtocounter{section}{1}
 \setcounter{equation}{0}
 \renewcommand{\thesection}{\Alph{section}}
 \section*{Appendix \thesection\protect\indent \parbox[t]{11.715cm} {#1}}
 \addcontentsline{toc}{section}{Appendix \thesection\ \ \ #1}
 }

\renewcommand{\thefootnote}{\fnsymbol{footnote}}

\newcommand{\be}{\begin{equation}}
\newcommand{\ee}{\end{equation}}
\newcommand{\ba}{\begin{aligned}}
\newcommand{\ea}{\end{aligned}}

\def\m1{\left(-1\right)^{F_i}}

\makeatletter
\def\sla@#1#2#3#4#5{{%
 \setbox\z@\hbox{$\m@th#4#5$}%
 \setbox\tw@\hbox{$\m@th#4#1$}%
 \dimen4\wd\ifdim\wd\z@<\wd\tw@\tw@\else\z@\fi
 \dimen@\ht\tw@
 \advance\dimen@-\dp\tw@
 \advance\dimen@-\ht\z@
 \advance\dimen@\dp\z@
 \divide\dimen@\tw@
 \advance\dimen@-#3\ht\tw@
 \advance\dimen@-#3\dp\tw@
 \dimen@ii#2\wd\z@  \raise-\dimen@\hbox to\dimen4{%
   \hss\kern\dimen@ii\box\tw@\kern-\dimen@ii\hss}%
 \llap{\hbox to\dimen4{\hss\box\z@\hss}}}}
\def\slashed#1{%
 \expandafter\ifx\csname sla@\string#1\endcsname\relax
   {\mathpalette{\sla@/00}{#1}}%
 \else
   \csname sla@\string#1\endcsname
 \fi}
\makeatother

\newcommand{\beq}{\begin{equation}}
\newcommand{\eeq}{\end{equation}}
\newcommand\beqa{\begin{eqnarray}}
\newcommand\eeqa{\end{eqnarray}}
\newcommand\bea{\begin{array}}
\newcommand\eea{\end{array}}

\newcommand{\nn}{\nonumber}
\newcommand{\neqa}{\nonumber\end{eqnarray}}
\newcommand{\la}{\label}

\newcommand{\color}[1]{}

\newcommand{\h}{\hat}
\renewcommand{\t}{\tilde}

\def\({\left(}
\def\){\right)}
\def\[{\left[}
\def\]{\right]}

\def\<{\langle}
\def\>{\rangle}

\def\sG{/\hspace{-0.25cm}G}

\begin{document}


\thispagestyle{empty}
\begin{flushright}\footnotesize
\texttt{LPTENS 08/38}\\
\texttt{SPhT-t08/NNN}\\
\vspace{2.1cm}
\end{flushright}

\renewcommand{\thefootnote}{\fnsymbol{footnote}}
\setcounter{footnote}{0}
\setcounter{figure}{0}
\begin{center}
{\Large\textbf{\mathversion{bold} The $AdS_4/CFT_3$ algebraic curve}\par}

\vspace{2.1cm}

\textrm{Nikolay Gromov$^{\alpha}$ and Pedro
Vieira$^{\beta}$}
\vspace{1.2cm}

\textit{$^{\alpha}$ Service de Physique Th\'eorique,
CNRS-URA 2306 C.E.A.-Saclay, F-91191 Gif-sur-Yvette, France;
Laboratoire de Physique Th\'eorique de
l'Ecole Normale Sup\'erieure et l'Universit\'e Paris-VI,
Paris, 75231, France;
St.Petersburg INP, Gatchina, 188 300, St.Petersburg, Russia } \\
\texttt{nikgromov@gmail.com}
\vspace{3mm}

\textit{$^{\beta}$ Laboratoire de Physique Th\'eorique
de l'Ecole Normale Sup\'erieure et l'Universit\'e Paris-VI, Paris,
75231, France;  Departamento de F\'\i sica e Centro de F\'\i sica do
Porto Faculdade de Ci\^encias da Universidade do Porto Rua do Campo
Alegre, 687, \,4169-007 Porto, Portugal} \\
\texttt{pedrogvieira@gmail.com}
\vspace{3mm}


\par\vspace{1cm}

\textbf{Abstract}\vspace{5mm}
\end{center}

\noindent
We present the $OSp(2,2|6)$ symmetric algebraic curve for the $AdS_4/CFT_3$ duality recently proposed in arXiv:0806.1218. It encodes all classical string solutions at strong t'Hooft coupling and the full two loop spectrum of long  single trace gauge invariant operators in the weak coupling regime. This construction can also be used to compute the complete superstring semi-classical spectrum around any classical solution. We exemplify our method on the BMN point-like string.
\vspace*{\fill}

\setcounter{page}{1}
\renewcommand{\thefootnote}{\arabic{footnote}}
\setcounter{footnote}{0}

\newpage



\section{Introduction and main results}
In \cite{MZ} integrability emerged once again \cite{MZ2} in the study of superconformal gauge theories. In this work Minahan and Zarembo wrote down a set of five Bethe equations yielding the complete $2$-loop spectrum of the three dimensional superconformal $SU(N)\times SU(N)$ Chern-Simons theory recently proposed by Aharony, Bergman, Jafferis and Maldacena in \cite{Aharony:2008ug} following \cite{Schwarz:2004yj}. This theory was conjectured in  \cite{Aharony:2008ug} to be the effective theory for a stack of M2 branes at a $Z_k$ orbifold point. In the large $N$ limit, the gravitational dual becomes M-theory on $AdS_4\times
S^7/Z_k$. For large $k$ and $N$ with
\beq
\lambda=N/k \equiv 8 g^2
\eeq
fixed, the dual theory becomes type IIA superstring theory in $AdS_4\times CP^3$. For subsequent interesting works see \cite{pos,BMN,MZ,today1,today2}.
In \cite{today1},\cite{today2} the superstring coset sigma model was constructed and shown to the classically integrable.

In this paper we present the algebraic curve construction for the $AdS_4/CFT_3$ duality at weak and strong coupling. The curves we present encode the full 2-loop spectrum of long single trace gauge invariant operators in the ABJM Chern Simons theory and the complete classical motion of free  type IIA superstring theory in $AdS_4\times CP^3$. The curve for the $AdS_5/CFT_4$ Maldacena duality was considered in \cite{C1,C2,C3,C4,C5,C6}.

For the string side we have a supercoset sigma model whose target space is
\beq
\frac{OSp(2,2|6)}{SO(3,1)\times SU(3)\times U(1)}
\eeq
which has $AdS_4\times CP^3$ as its bosonic part. The algebraic curves allows us to map each classical solution to a corresponding Riemann surface which encodes an infinite set of conserved charges particular to the classical solution in study. The map goes as follows. Given a classical solution we can diagonalize the monodromy matrix
\beq
\Omega(x)=
P\exp \int d\sigma J_\sigma (x)  \la{monodromyIntro}
\eeq
where $J(x)$ is the flat connection, present for integrable theories and computed for the model in study in \cite{today1,today2} and for the $AdS_5\times S^5$ superstrings in \cite{BPR}. The eigenvalues of such matrix (as of any generic matrix) live in a Riemann surface whose size is roughly speaking the size of the matrix\footnote{For some matrices, such as for example elements of $SO(2N+1)$, some eigenvalues might be trivial.}. In our case, as explained bellow, we will see that the logarithms of these eigenvalues can be organized into a 10-sheeted Riemann surface whose properties are listed bellow.

Turning the logic around, the algebraic curve construction allows one to trade the study of the intricate non-linear equations of motion by the construction of Riemann surfaces with some precise prescribed analytical properties. The string dynamics can be translated to the study of analytical properties of algebraic curves, a well developed subject in algebraic geometry.

To illustrate what we mean let us describe the algebraic curve studied in this paper. To find the complete classical spectrum of the theory we should proceed as follows:
We should build ten-sheeted Riemann surfaces whose branches\footnote{Actually as will become clear latter the quasi-momenta define an infinite genus curve and to obtain a ten-sheeted Riemann surface we should take for example the derivative of this quasimomenta w.r.t. $x$.}, called quasi-momenta, depend on a spectral parameter $x \in \mathbb{C}$ and are denoted by $\{ q_1,  q_2,   q_3,   q_4,   q_5,   q_6,  q_7,  q_8,q_9,  q_{10}\}$. They are not independent but rather $\{    q_1,     q_2,   q_3,   q_4,   q_5\}=\{-    q_{10},-    q_9,-  q_8 ,-  q_7,-   q_6\}$.
\begin{figure}[t]
\centering \resizebox{95mm}{!}{\includegraphics{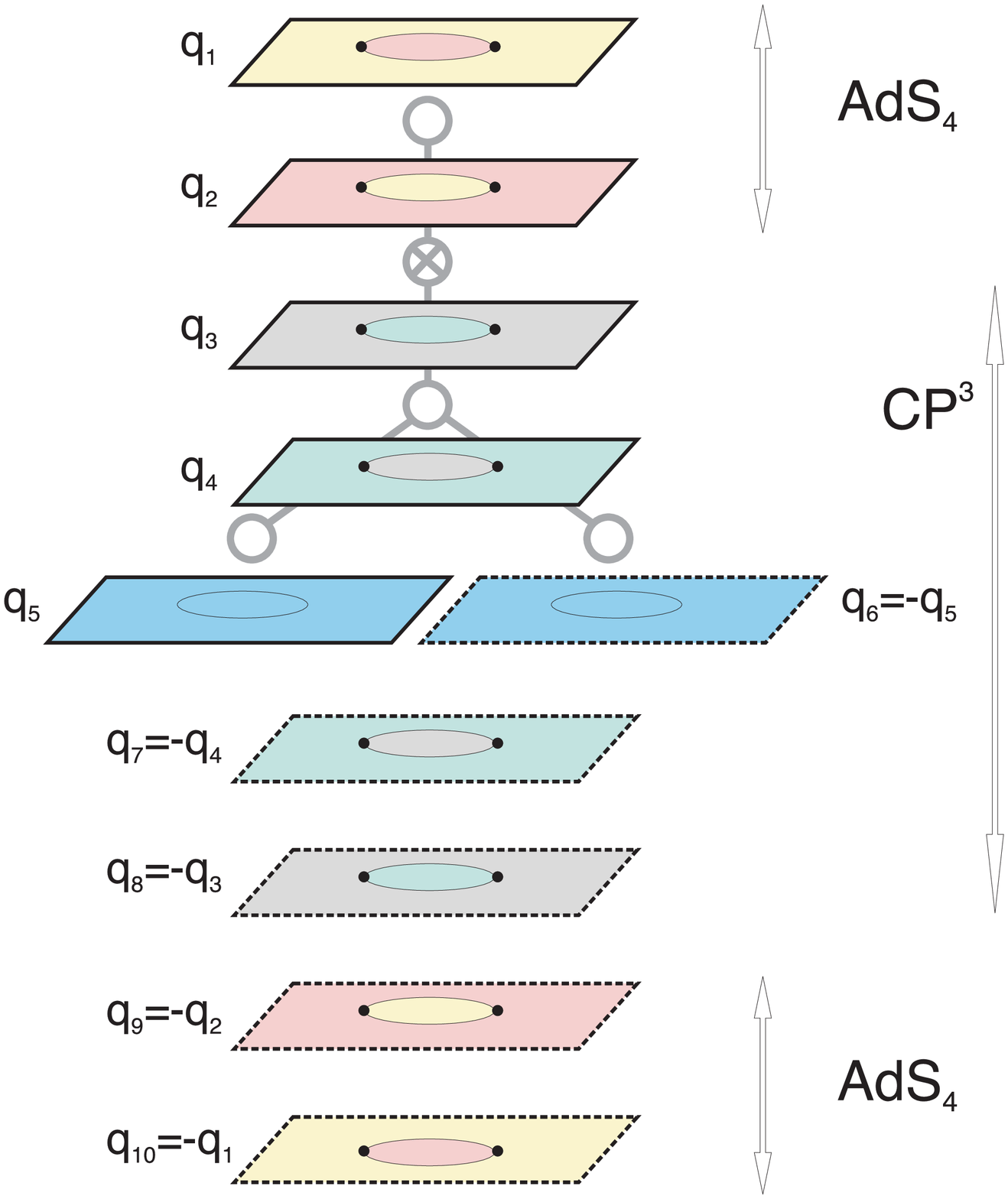}}
\caption{\small Full $AdS_4 \times CP^3$ algebraic curve in the
$ \textbf{10}$ representation. Poles uniting $AdS_4$ and $CP^3$ quasimomenta
are fermionic excitations. Regions which are trivially related are
painted with the same colour. Poles at $x=\pm 1$ are marked by black
filled circles. The $OSP(2,2|6)$ emerges naturally and notice that the
black dots disappear as we jump though the dynkin nodes
whose Dynkin labels are non-zero, precisely as expected -- Bethe
equations should be difference of quasimomenta and only therefore this pattern reflects the $SU(N) \times SU(N)$ staggered spin chain of Minahan and Zarembo \cite{MZ}.} \label{curve}
\end{figure}
These Riemann surfaces must obey the following analyticity properties:
\begin{enumerate}
\item Generically square root cuts may connect different pairs of sheets. When going through each cut the quasi-momenta might gain a multiple integer of $2\pi$,
\beq
q_i^+-q_j^-=2\pi n_{ij} \,\,\, , \,\,\, x\in \mathcal{C}_{ij} \la{qij}
\eeq
where the superscript $\pm$ indicated the function is evaluated immediately above/below the square root cut. The set of integers $\{n_{ij}\}$ characterize the several cuts of the Riemann surface, i.e. they are are moduli of the algebraic curve.
\item
Each cut is also parametrized by a filling fraction
\beqa
S_{ij}&=&\frac{g}{\pi i} \oint_{\mathcal{C}_{ij}} dx
\(1-\frac{1}{x^2}\) q_i(x)
\eeqa
which roughly speaking measures how big the cut is. (From the point of view of the classical solutions these are the action variables.)
\item The quasi-momenta must behave as
\beqa
\(\begin{array}{l}
   q_1(x)  \\
   q_2(x)  \\
 q_3(x)  \\
 q_4(x)  \\
 q_5(x)  \end{array}\)=-\(\begin{array}{l}
   q_{10}(x) \\
   q_9(x) \\
 q_8(x) \\
 q_7(x) \\
 q_6(x) \end{array}\)\simeq \frac{1/2}{x\pm 1}\( \begin{array}{c}
\alpha_\pm \\
\alpha_\pm \\
\alpha_\pm \\
\alpha_\pm \\
0
\end{array}\) \la{pq1}
\eeqa
close to the singular points $x=\pm 1$. The constant $\alpha_\pm$ has no significance from the target space point of view, the only think we should keep in mind is that the residues must be synchronized (Physically this is a manifestation of the  Virasoro constraints imposed on the classical solutions).
\item
The curve should possess the inversion symmetry
\beqa
\(\begin{array}{l}
   q_1(1/x)  \\
   q_2(1/x)  \\
 q_3(1/x)  \\
 q_4(1/x)  \\
 q_5(1/x)  \end{array}\) =\(\begin{array}{c}
0\\
0 \\
2\pi m \\
2\pi m \\
0 \end{array}\)+\(\begin{array}{l}
-    q_{2}(x) \\
-    q_1(x) \\
-  q_4(x) \\
-  q_3(x) \\
+  q_5 (x)\end{array}\)=\(\begin{array}{c}
0\\
0 \\
2\pi m \\
2\pi m \\
0 \end{array}\)+
\(\begin{array}{l}
+    q_{9} (x)\\
+    q_{10} (x)\\
+  q_7 (x)\\
+  q_8 (x)\\
-  q_6(x) \end{array}\) \la{pq2}
\eeqa
with $m$ being an integer.
\item Finally the large $x$ asymptotics of the Riemann surface read\footnote{The state labeled by
$(M_u,M_r,M_v)$ belongs to the $SU(4)$ representation with Dynkin labels $[L-2M_u+M_r,M_u+M_v-2M_r,L-2M_v+M_r]$}
\beq
\(\begin{array}{l}
    q_1 (x) \\
    q_2  (x)\\
  q_3  (x)\\
  q_4  (x)\\
  q_5 (x) \end{array}\)=\(\begin{array}{l}
    q_{10} (x)\\
    q_9 (x)\\
  q_8 (x)\\
  q_7 (x)\\
  q_6(x) \end{array}\)\simeq
\frac{1}{2g x}\(
\begin{array}{l}
 L+E+S \\
L+E-S \\
L-M_r \\
L+M_r-M_u-M_v\\
M_v-M_u \\
\end{array}\)\,. \la{pq3}
 \eeq
\end{enumerate}
Therefore, if we enumerate all possible Riemann surfaces with the properties just listed, then, it suffices to evaluate the quasimomenta at large values of the spectral parameter to obtain the energy spectrum of all classical string solutions as a function of the global charges charges and several moduli of the algebraic curve.

Notice that each cut of the algebraic curve is characterized by a discrete label $(i,j)$, corresponding to
the two sheets being united, an integer $n$, the multiple of $2\pi$
mentioned above, and a real filling fraction. These three quantities are
the analogues of the polarization, mode number and amplitude of the
flat space fourier decomposition of a given classical solution.

Then we have a clear geometrical picture of semi-classical quantization in the context of these algebraic curves. Namely a classical solution will be represented as some algebraic curve with same large cuts uniting several pairs of sheets. Quantum fluctuations correspond to adding small singularities -- microscopic cuts or poles -- to this Riemann surface \cite{GV1}. The different choices of sheets to be connected in this way correspond to the different string polarizations we can excite. We will exemplify this procedure on the example of the simplest classical solution -- the BMN \cite{BMNreal} string studied in the present framework in \cite{BMN}. This method can be easily generalized to more complicated solutions and to the study of the ground state energy around any classical solution \cite{applications}.

One needs to carry on the investigation of integrability in $AdS/CFT$, so successful for the most famous $AdS_5/CFT_4$ duality \cite{huge}. The results are of course unpredictable but one thing can be taken for granted -- the understanding of the integrable structures behind all these beautiful theories can only  deepen our understanding of non-perturbative gauge theories and theories of quantum gravity.
\section{String algebraic curve}
In this paper we analyze the algebraic curve for free type IIA
superstrings in $AdS_4/CP^3$.
To compute the algebraic curve we could use the flat connection in
\cite{today1,today2} which has superficially the same form as that
found by Benna Polchinki and Roiban for the $AdS_5\times S^5$ strings
\cite{BPR}. Armed with the experience of what happens in the $AdS_5/CFT_4$ duality
\cite{C1,C6,GV1} we follow a shortcut. We shall use the purely bosonic part
of the action to compute the $CP^3$ and the $AdS_4$ algebraic curves.
Only Virasoro will couple them. Then, to lift it from the classical bosonic curve to the complete semi-classical curve for the full super
group, we will simply allow the several sheets of the two curves to be
connected by further small cuts or poles.
If the poles connect $CP^3$ sheets with $AdS_4$ sheets then we will be studying the
missing fermionic excitations. Of course, if we would proceed as in
\cite{C6} using the full flat connection in \cite{today1,today2} we would
find exactly the same results as can be easily checked.

Technically, our  treatment is very similar to the one in \cite{C3} where the $SO(6)$ bosonic string was studied and the generalization to $SO(2n)$ was carried on. This is not surprising since $OSp(2,2|6)$ is not very different from $SO(10)$.

\subsection{Bosonic flat connection}
The bosonic part of the $AdS_4/CP^3$ type IIA free superstring theory reads
\beq
S=\sqrt {2\lambda} \int d\sigma d\tau \( \mathcal{L}_{CP^3}
+\mathcal{L}_{AdS^4} \)
\eeq
where
\beq
\mathcal{L}_{AdS^4}= -\frac{1}{4}\(\partial_\mu n \cdot \partial_\mu n - \Lambda
\(n\cdot n-1\)\) \,,
\eeq
and
\beq
\mathcal{L}_{CP^3}= \(D_\mu z \)^\dagger \cdot D_{\mu} z-\Lambda'
\(z^\dagger\cdot z-1\) \,.
\eeq
Here $n$ and $z$ are vectors made out of the  embedding coordinates of
the anti de-Sitter and the projective space. Thus
\beq
n=(n_1,\dots,n_5) \,\, , \,\, n\cdot n=n_1^2+n_2^2-n_3^2-n_4^2-n_5^2
\eeq
with $n_i$ real while
\beq
z=(z^1,\dots,z^4) \,\, , \,\, z^\dagger\cdot z=|z^1|^2+\dots+|z^4|^2
\eeq
where $z^I$ are complex numbers. In what follows whenever the index
structure is obvious we omit it. The $z^I$ are also identified up
to a phase, $z^I \simeq e^{i\varphi} z^I$. This $U(1)$ gauge symmetry
is accounted by the gauge field $A_\mu$ appearing in
\beq
D_\mu z=\partial_\mu z + i A_\mu z\,.
\eeq
The equations of motion for the connection yield $z \cdot \(D_\mu
z\)^\dagger-\(D_\mu z\)\cdot z^\dagger=0$ while the constrain
$\partial_\mu(z^\dagger\cdot z)=0$ yields $z \cdot \(D_\mu
z\)^\dagger+\(D_\mu z\)\cdot z^\dagger=0$ and therefore, on-shell, we
have separately
\beq
\(D_\mu z\)\cdot z^\dagger=z\cdot \(D_\mu z^\dagger\)=0 \,. \la{onshell}
\eeq
Analogously, for the $n$ field we have
\beq
n\cdot \(\partial_\mu n\)=0\,. \la{onshell2}
\eeq
Next it is useful to introduce the element
\beq
h=\(\begin{array}{c|c}
1-2\,z^\dagger \otimes z &   \\ \hline
&  1-2\,n \otimes n
\end{array} \) \,\,\, \Leftrightarrow h_{AB}= \(\begin{array}{c|c}
\delta_I^J-2\,z^\dagger_I z^J  &   \\ \hline
&  \delta_{ij}-2\,n_i n_j
\end{array} \) \,, \la{h}
\eeq
and the connection
\beq
j=h^{-1} d h \,. \la{jdef}
\eeq
It is easy to see that when $z^\dagger\cdot z=n\cdot n=1$ and
(\ref{onshell}), (\ref{onshell2}) hold we have
\beq
j_{AB}=\(\begin{array}{c|c}
 j_{\rm AdS} &   \\ \hline
& j_{\rm CP} \eea \)=2  \(\begin{array}{c|c}
  n_i \(\partial_\mu n_j\)-\(\partial_\mu n_i \) n_j &   \\ \hline
& z^\dagger_I \(D_\mu z\)^J - \(D_\mu z\)^\dagger_I  z^J
\end{array} \)    \la{jblocks}
\eeq
and the action becomes
\beq
S=-\frac{g}{4} \int d\sigma d\tau \,{\rm STr}_{\rm OSp} \(j_\mu^2\) \,.
\eeq
where\footnote{The reason for this definition lies in the relations between quadratic
Casimirs for $SO(5),Sp(4)$ and $SU(4),SO(6)$
$$
{\rm Tr}_{SO(5)}(j^2)=2{\rm Tr}_{Sp(4)}(j^2)\;\;,\;\;
{\rm Tr}_{SU(4)}(j^2)=\frac{1}{2}{\rm Tr}_{SO(6)}(j^2)
$$}
\beq
{\rm STr}_{\rm OSp} \(j^2\)\equiv -\frac{1}{2}{\rm Tr}\(j_{\rm AdS}^2\)+2{\rm Tr}\(j_{{\rm CP}}^2\)
\eeq
Note also that the Virasoro constraint now implies
\beq
{\rm STr}_{\rm OSp} \(j_1\pm j_0\)^2 = 0\,.
\eeq
At this point, we have the flatness condition
\beq
dj+j\wedge j=0   \,,
\eeq
following from the expression (\ref{jdef}) of the connection, and
\beq
d*j=0\,,
\eeq
encoding the equations of motion for both the $n$ and the $z$ field.
These two equations follow from the flatness condition for the Lax
connection
\beq
J(x)= \frac{j+x *j}{1-x^2} \la{Jx}
\eeq
as can be easily checked by collecting powers of $x$. The new variable
$x$ appearing in (\ref{Jx}) is a completely arbitrary complex number called spectral
parameter.
Using this flat connection we can build the monodromy matrix
\beq
\Omega(x)=
P\exp \int d\sigma J_\sigma (x)  \la{monodromy}
\eeq
where we integrate over a constant $\tau$ worldsheet loop.
At this point integrability comes into stage. The connection being
flat, the eigenvalues of the monodromy matrix are independent of
$\tau$ and since moreover they depend on a generic complex number
$x$ they define an infinite set of conserved charges. For
example, each coefficient in the taylor expansion of the eigenvalues around a particular point $x^*$ is a conserved charge.  The existence of this large number
of conserved charges render the sigma model (at least classically) integrable.

We want to study the algebraic curve construction for this integrable
model. This will map each classical string solution to a Riemann
surface with precise analytical properties. The study of classical
solutions in $AdS_4\times CP^3$ can then be reduced to the problem of
making a catalogue of all Riemann surfaces compatible with the
prescribed analytical properties.

\subsection{The $AdS_4\times CP^3$ algebraic curve}
In this section we study the eigenvalues of the monodromy matrix
(\ref{monodromy}). We will first consider purely bosonic solutions and work out
the full supercurve in the next section.
From the form of the flat connection, in particular from the fact that
each of the blocks in (\ref{jblocks}) is manifestly traceless, we find
that the connection is explicitly block diagonal and thus its
eigenvalues will split into two groups: The eigenvalues coming from
the $CP^3$ part
\beq
\{ e^{i \t p_1},\dots, e^{i \t p_4} \} \,,
\eeq
with
\beq
\t p_1(x)+\dots+\t p_4(x)=0 \la{tr1}\,,
\eeq
and those coming from the diagonalization of the $AdS$ block,
\beq
\{ e^{i \hat p_1},\dots, e^{i \h p_4}, 1 \} \,,
\eeq
where moreover
\beq
\hat p_{3}(x)+\hat p_2(x)=0 = \h p_4(x)+\h p_1(x) \,. \la{tr2}
\eeq
To find the eigenvalues of the monodromy matrix we solve a polynomial
characteristic equation. This defines a algebraic curve for the
eigenvalues $\lambda$. Thus, the eigenvalues can the thought of as
different branches of the same Riemann surfaces with square root cuts
uniting the several sheets. For example when crossing a cut
$\mathcal{C}$ shared by the eigenvalues $e^{i\h p_1}$ and $e^{i
\h p_2}$ we simply change Riemann sheet,
\beq
\(e^{i \h p_2}\)^+-\(e^{i \h p_1}\)^- = 0 \,\,\, , \,\,\, x \in
\mathcal{C} \la{expp23pre}
\eeq
where the superscript $\pm$ indicates the function is evaluated
immediately above/below the cut. The quasi-momenta on the other hand
are not exactly the eigenvalues but rather the logarithms of the
eigenvalues. Thus when crossing the very same cut the quasimomenta
will in general also gain an integer multiple of $2\pi $,
\beq
 \h p_2^+- \h p_1^- = 2\pi n  \,\,\, , \,\,\, x \in \mathcal{C} \,. \la{p23pre}
\eeq
In general we can have several cuts uniting different pairs of sheets and
\beq
p_i^+-p_j^-=2\pi  n \,\,\, , \,\,\, x \in \mathcal{C}_{ij}\,,
\eeq
on each cut connecting two quasimomenta $p_i$ and $p_j$. To
parametrize each cut we also introduce the usual filling fractions
\beqa
\hat S_{ij}&=&\frac{g}{2\pi i} \oint_{\mathcal{C}_{ij}} dx
\(1-\frac{1}{x^2}\) \hat p_i(x)\;\;,\;\;\t S_{ij}=\frac{g}{\pi i} \oint_{\mathcal{C}_{ij}} dx
\(1-\frac{1}{x^2}\) \t p_i(x)
\eeqa
for the cuts uniting $p_i$ and $p_j$.   Each cut of the algebraic
curve is characterized by a discrete label $(i,j)$, corresponding to
the two sheets being united, an integer $n$, the multiple of $2\pi$
mention above, and a real filling fraction. These three quantities are
the analogues of the polarization, mode number and amplitude of the
flat space fourier decomposition of a given classical solution.

The study of the analytical properties of the quasi-momenta follows
closely the analysis done in the context of the $AdS_5/CFT_4$ duality
in \cite{C6}. Let us enumerate all these properties and then explain their origin.

For large values of the spectral parameter, the quasimomenta behave as
\beqa
\(\h p_1,\h p_2 ,\h p_3 ,\h p_4\) &\simeq & \frac{1}{gx}\(L+E,S,-S,-L-E\)  \, ,\\ \nn
\(\t p_1,\t p_2,\t p_3,\t p_4\)& \simeq &
\frac{1}{2gx}\(L-M_u,M_u-M_r,M_r-M_v,-L+M_v\)  \, ,
\eeqa
for a state belonging to the $SU(4)$ representation
with Dynkin labels $[L-2M_u+M_r,M_u+M_v-2M_r,L-2M_v+M_r]$ (which
should be positive).

There are two simple poles at $x=\pm  1$ which are synchronized between the $AdS_4$ and the $CP^3$ quasi-momenta,
\beq
\(\h p_1,\h p_2 ,\h p_3 ,\h p_4;\t p_1,\t p_2,\t p_3,\t p_4\) \simeq \frac{1}{x\pm 1}\( \alpha_\pm,0,0,-\alpha_\pm;\frac{\alpha_\pm}{2},0,0,-\frac{\alpha_\pm}{2}\)  \, , \la{xpm1}
\eeq
and finally the algebraic surface exhibits a $x\to 1/x$ inversion symmetry under which
\beq
\begin{array}{l}
\h p_1(1/x)=-\h p_1(x) \\
\h p_2(1/x)=+\h p_2(x) \,\\
\h p_3(1/x)=+\h p_3(x) \,\\
\h p_4(1/x)=-\h p_4(x)  \, \end{array}
\,\,\, , \,\,\,
\begin{array}{l}
\t p_1(1/x)=\t p_4(x)+2\pi m\, \,\\
\t p_2(1/x)=\t p_2(x) \,\\
\t p_3(1/x)=\t p_3(x) \,\\
\t p_4(1/x)=\t p_1(x)-2\pi m \, \end{array}
 \la{inversion}  \,.
\eeq
with $m$ being an integer dependent on the classical solution to which these quasi-momenta are associated.

Let us now briefly explain the origin of these analytical properties. The fact that the quasi-momenta encode the global charges of the
classical solutions at the $x \to \infty$ asymptotics follows from  the large $x$ behavior of the monodromy matrix,
\beq
\Omega(x) \simeq
1+   \frac{1}{x}\int d\sigma j_\tau \,.
\eeq
From the form of the flat connection we see that in general
the quasi-momenta can have simple poles at $x=\pm 1$. The reason why
only four of them -- two in $CP^3$ and two in $AdS_4$  -- have
non-vanishing residues follows from the very
particular form of the flat connection.
For example for $x\simeq 1$ we have  $J(x) \propto j_+$ and thus
\beq
j_+ \cdot v=0
\eeq
if $v$ is orthogonal to both  $z_I^\dagger$, $D_+ z_I^\dagger$, $n_i$
and $\partial_+ n_i$. Following the arguments in \cite{C3}, this can
be shown to imply that only two $CP^3$ and two $AdS_4$ quasimomenta
have poles. Since moreover (\ref{tr1}) and (\ref{tr2}) we
immediately see that the residues at these poles must be symmetric. Moreover
the Virasoro constraints ${\rm STr}(j_\mu^2)=0$ synchronizes the poles in the anti de-Sitter
and projective space as in (\ref{xpm1}) (exactly as in \cite{C5}).

Next we notice that $h^{-1}=h$. This has important consequences for
the algebraic curve. It implies that
\beq
\Omega(x)= h^{-1}(2\pi) \Omega(1/x) h(0) \la{Omegaxto1/x}
\eeq
and therefore the eigenvalues of the monodromy matrix associated to
some  closed string classical solution are at most exchanged between
themselves under the inversion map $x \to 1/x$. The precise way in
which the quasi-momenta are exchanged is in general a subtle business
\cite{C3,C5}. For example, a priori from (\ref{Omegaxto1/x}) it
seems that we could not infer that $\t p_2(1/x)$ is not exchanged with
$\t p_3(x)$ for example. The reason why this can not happen and the
inversion symmetry we postulated is OK is the following: There are
solutions with $z_1,z_2,z_3\neq 0$ but $z_4=0$. For those solutions the last line and column of
the current $J(x)$ is made out of zeros. Thus $\Omega(x)$ will have one eigenvalue exactly equal to $1$.
In other words, one of the quasimomenta is strictly zero.
Since $\t p_1$ and $\t p_4$ have poles this quasimomenta must be either
$\t p_2$ or $\t p_3$. But then, if for example $\t p_3=0$ while the other three
quasimomenta are nontrivial vanishing then, clearly, $p_2(1/x)\neq p_3(x)$! In the
same way we can justify the remaining relations in (\ref{inversion})\footnote{As we mentioned in the beginning we could also have constructed the curve using the flat connection in \cite{today1,today2}.
The $x\to 1/x$ symmetry we just discussed appears in these works as a consequence of the $\mathbb{Z}_4$ grading of the superalgebra (see equation at the end of section 4.1 in \cite{today2}). In the context of the $AdS_5\times CFT_4$ correspondence -- see \cite{C5} -- this was also the case. Had we used the flat connections in these works and we would have found the same inversion symmetry properties.}.

\subsection{Full algebraic supercurve}
In this section we generalize the classical bosonic algebraic curve described in the previous section to the semi-classical and supersymmetric  $OSp(2,2|6)$ algebraic curve. The smallest representation of this symmetry group -- which behaves in many aspects as $SO(10)$ -- is
$\textbf{10}$ dimensional so we should find a nice linear combination of the quasimomenta in the previous section yielding $\textbf{10}$ functions describing a single algebraic curve with manifest $OSp(2,2|6)$ symmetry. Then, to include fermions, we simply allow  for extra poles between the $AdS_4$ and the $CP^3$ quasimomenta! A proper linear
combination is the following reorganization of the
quasimomenta into a set of ten functions
\beqa
\{ q_1,q_2,q_3,q_4,q_5\}=\,\left\{\frac{\h p_1+\h p_2}{2}, \frac{\h p_1-\h p_2}{2}, \t p_1+\t p_2, \t p_1+\t p_3,\t p_1+\t p_4\right\}  \,, \la{qs1}
\eeqa
and
\beqa
\{ q_6,q_7,q_8,q_9,q_{10}\}=-\,\{ q_1,q_2,q_3,q_4,q_5\}  \,. \la{qs2}
\eeqa
\begin{figure}[t]
\centering \resizebox{80mm}{!}{\includegraphics{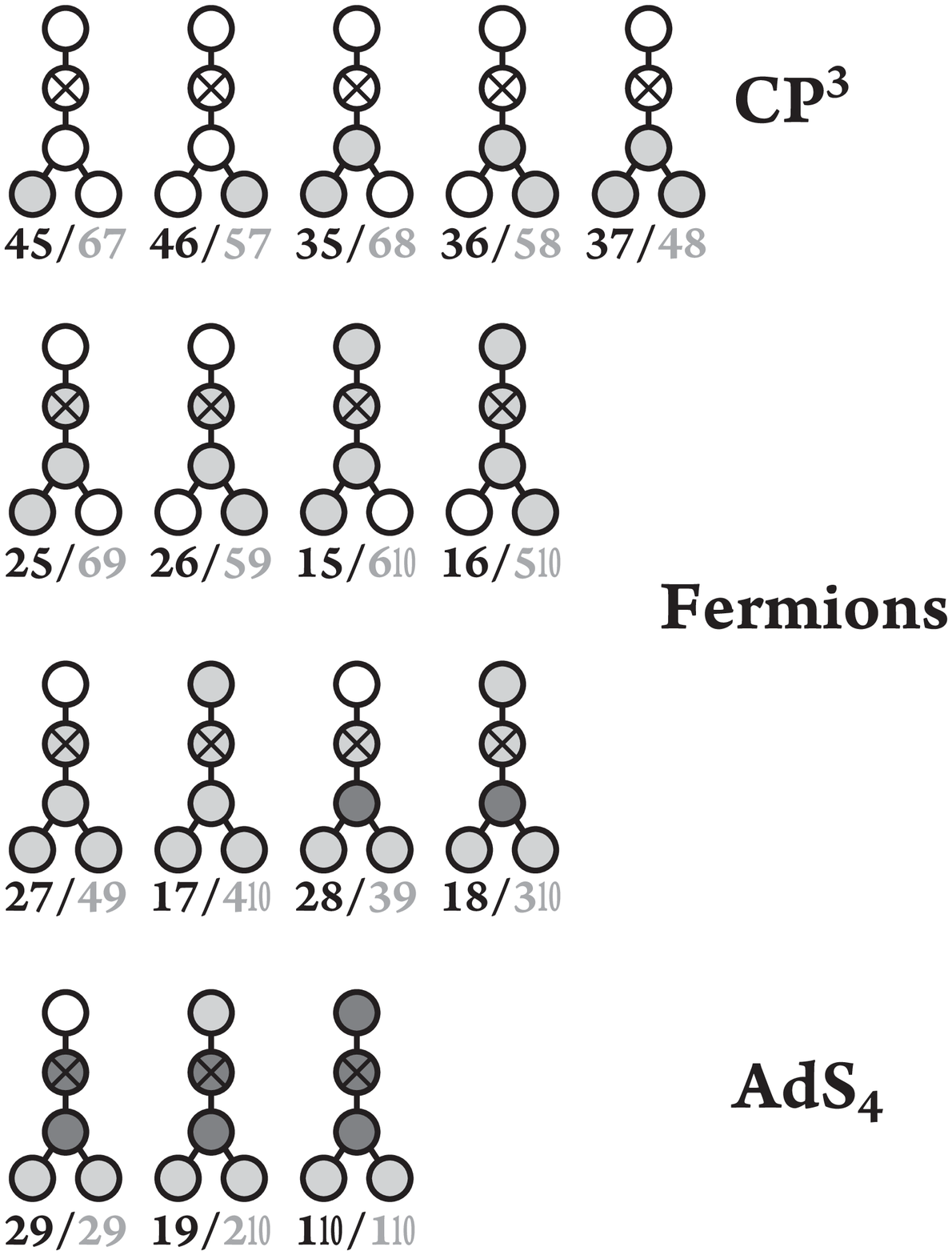}}
\caption{\small The several states in the Hilbert space can be constructed in the usual oscillator representation. There is one oscillator per Dynkin node of the $OSp(2,2|6)$ super Dynkin diagram. A light (dark) gray shaded node corresponds to an oscillator excited once (twice). From the Chern-Simons Bethe ansatz point of view, the number of times each oscillator is excited is the same as the number of Bethe roots of the corresponding type. Thus, for example, in the notation of \cite{MZ}, the last fermionic excitation corresponds to a bound state of one root of each type $u,v,w,s$ and two Bethe roots $r$. From the string point of view fluctuations correspond to poles uniting the several sheets of the algebraic curve. Close to each fluctuation we represented some numbers like $45/67$ for the first fluctuation. They indicate which momenta are being united by this pole. In this case it is momenta $q_4$ and $q_5$. Since (\ref{qs2}) automatically $q_6$ and $q_7$ also share a pole.} \label{excitations}
\end{figure}
These ten functions can be though of as the several sheets of a single function taking values in a ten-sheeted Riemann
surface as represented in figure \ref{curve}. Notice that they organize in a nice explicitly $OSp(2,2|6)$ symmetric way.
From the properties derived in the previous section the properties for the $q_i$ listed in the introduction follow. In particular we have (\ref{pq1}), (\ref{pq2}) and (\ref{pq3}).

Next, to understand the quasi-classical quantization of any classical solution we add extra pole singularities to the different pairs of sheets of the algebraic curve associated with the solution we want to quantize. The several pairs of sheets to be connected in figure \ref{curve}  correspond to the different physical polarizations for the quantum fluctuations. Thus, we are in need of a map between the several possible excitations of the string Hilbert space (or of the dual gauge theory) and the several pairs or Riemann surfaces.

This map is provided by figure \ref{excitations} where we listed all $16=8+8$ physical excitations. The fluctuations are identified by the corresponding excitations of the $OSp(2,2|6)$ Dynkin diagram with Dynkin labels as in \cite{MZ}. Since the asymptotics of the curve can also be related to the Dynkin labels of a given state this suffices to identify which pairs of sheets are connected for each quantum fluctuation. See  \cite{C5,GV1} for similar analysis  in the context of the $AdS_5/CFT_4$ duality.


In the next section we will explicitly apply figure \ref{excitations} to the semi-classical quantization of the BMN string.

\subsection{BMN string}
The BMN point-like string has $z_1=\frac{1}{\sqrt{2}}e^{i\omega
\tau/2}$, $z_2=\frac{1}{\sqrt{2}} e^{-i\omega \tau/2}$ and $n_1+i n_2=
e^{i \omega \tau}$. Computing the charges of this solution we find
\beq
L=4\pi g\,\omega =\pi \sqrt{2 \lambda}\,\omega\;\;,\;\;E=0 \, \la{omegaJ}
\eeq
to check that one can use that the $AdS_5$ time is given by $-i\log(n_1+in_2)$.
To compare the results we will find below with those in \cite{BMN} we notice that

\beq
\frac{n^2}{\omega^2}= \frac{2 \pi ^2 \lambda \,n^2}{L^2}\,.
\eeq
We now plug the embedding coordinates into (\ref{jblocks}) and compute
the path ordered exponential in (\ref{monodromy}). Since the string is
point-like there is no $\sigma$ dependence and this computation is
trivial. We can then compute the eigenvalues of the monodromy matrix
(\ref{monodromy}) and from them we find the $\bf{10}$ $q_i$'s using
(\ref{qs1}) and (\ref{qs2}). We obtain
\beq
q_{1,\dots,4}=-q_{6,\dots,10}=\frac{2\pi \omega x }{x^2-1}  \la{list} \,,
\eeq
and
\beq
q_{5,6}=0 \, . \la{zeros}
\eeq
The BMN string is the simplest possible algebraic curve. It is in fact
the \textit{vacuum curve}, all sheets are empty except for the two
single poles at $x=\pm 1$.

We can now exemplify the computation of the quasi-classical spectrum
in the algebraic curve language and reproduce the recent results of
\cite{BMN}.
The $16$ physical excitations are represented in figure
\ref{excitations}. Notice that the first four $CP^3$ fluctuations  and
the last four fermionic fluctuations corresponds to poles shared by a
quasimomenta in the list (\ref{list}) with one of the two quasimomenta
in (\ref{zeros}). The position of these fluctuations is given by
\cite{GV1}
\beq
q_i(x_n)-q_j(x_n)=2\pi n \la{general} \,.
\eeq
The integer $n$ is the generalization of the Fourier mode in flat
space -- it is meaningful around any classical solution, no matter how
non-linear and non-trivial this solution might be. For the
fluctuations we are discussing this equation reads
\beq
\frac{2\pi \omega x_n}{x_n^2-1} =2\pi n  \,,
\eeq
and we should pick the solution in the physical region $|x|>1$. On the
other hand all the remaining eight fluctuations connect two
quasimomenta in (\ref{list}). Therefore, from (\ref{general}), we will
find that the position of these fluctuations is fixed by
\beq
\frac{2\pi \omega x_n}{x_n^2-1} =\pi n  \,.
\eeq
So the position of half of the fluctuations is the same as the
position of the other half with doubled mode number. This already
points towards the structure of the fluctuation energies observed in
\cite{BMN}. We also recall that a fluctuation pole at position $y$
should have a residue \cite{GV1}
\beq
\alpha(y)=\frac{1}{2g} \frac{y^2}{y^2-1} \la{alpha}
\eeq
We will now consider separately the $CP^3$, $AdS_4$ and Fermionic
excitations to understand how to compute the fluctuation energies
around a classical solution in the algebraic curve formalism. The
computations are conceptually as in \cite{GV1} so we will simply
present the results for the perturbed quasi-momenta with very few
explanations.

A technical detail: When computing the fluctuation spectrum we will
add always a fluctuation with mode number $n$ and another with mode
number $-n$ to keep the string total world-sheet momentum zero in the
process. We could alternatively excite all polarizations at the same
time while obeying the level matching condition
\beq
\sum_{ij,n} n N^{ij}_n=0\,,
\eeq
with $N^{ij}_n$ being the number of fluctuations with polarization
$(i,j)$ and mode number $n$. This would lead to the same results but
with cluster our expression so we will chose to add always a single
pair of fluctuations at a time.

\subsubsection{$CP^3$ excitations}
There are two types of fluctuations in $CP^3$: The first four in
figure \ref{excitations} and the fifth one. The former corresponds to
a pole connecting a quasi-momenta in (\ref{list}) with an empty one in
(\ref{zeros}) whereas the latter corresponds to a pole shared by $q_3$
and $q_7$, both in (\ref{list}).  Let us consider one of the
fluctuations of the first type, say the first one in figure
\ref{excitations}. For this fluctuation
\beqa
&&\delta   q_3=-\delta  q_{8}=  +\sum_{\pm }
\frac{\alpha(1/x)}{1/x-x_{\pm n}} \\
&&\delta  q_4=-\delta q_{7}=  -\sum_{\pm } \frac{\alpha(x)}{x-x_{\pm n}} \\
&&\delta  q_5=-\delta  q_{6} =  +\sum_{\pm } \frac{\alpha(x)}{x-x_{\pm
n}}+ \sum_{\pm } \frac{\alpha(1/x)}{1/x-x_{\pm n}}\\
&&\delta  q_{1,2}=-\delta  q_{9,10} =+\frac{\alpha(x)\,\delta E}{x}\,.
\eeqa
so that from the synchronization of poles at $x=\pm 1$ we find
\beqa
\delta E = \sum_{\pm} \frac{1}{x_{\pm n}^2-1}=\sum_{\pm n}
\sqrt{\frac{1}{4}+\frac{n^2}{\omega^2}}-\frac{1}{2}\,.
\eeqa
The fifth fluctuation in $CP^3$ connects $q_3$ and $q_7$ (and
therefore automatically at $q_4=-q_7$ and $q_8=-q_3$) so we have
\beqa
&&\delta q_4=-\delta  q_{7}=  -\sum_{\pm } \frac{\alpha(x)}{x-x_{\pm
n}}+ \sum_{\pm } \frac{\alpha(1/x)}{1/x-x_{\pm n}} \\
&&\delta   q_3=-\delta   q_{8}=  -\sum_{\pm }
\frac{\alpha(x)}{x-x_{\pm n}}+ \sum_{\pm }
\frac{\alpha(1/x)}{1/x-x_{\pm n}} \\
&&\delta   q_{1,2}=-\delta   q_{9,10} =+  \frac{\alpha(x)\,\delta E}{x}\,.
\eeqa
and we find in this case
\beqa
\delta E= \sum_{\pm} \frac{2}{x_{\pm n}^2-1}= \sum_{\pm
n}\sqrt{1+\frac{n^2}{\omega^2}}-1 \,.
\eeqa
These are precisely the results of \cite{BMN}.
\subsubsection{$AdS^4$ excitations}
Here we must be careful. The first and third $AdS^4$ fluctuations have
two excitations in the last Dynkin node -- see figure
\ref{excitations}. This means that for those we should double the
residue (\ref{alpha}). Let us consider the last one for illustration
(for the first one we get the same result of course). We have
\beqa
&&\delta   q_1=-\delta  q_{10}=  + \sum_{\pm }
\frac{2\alpha(x_n)}{x-x_{\pm n}} \\
&&\delta   q_2=-\delta  q_{9} = -\sum_{\pm } \frac{2\alpha(x_n)}{1/x-x_{\pm n}}
\eeqa
and thus, from the large $x$ asymptotics,
\beqa
\delta E= \sum_{n=\pm} \frac{x_n^2+1}{x_n^2-1}=\sum_{\pm n}
\sqrt{1+\frac{n^2}{\omega^2}} \la{w1}\,.
\eeqa
As for the middle fluctuation in figure \ref{excitations}, we have
\beqa
\delta   q_1=-\delta q_{10}=  + \sum_{\pm }
\frac{\alpha(x_n)}{x-x_{\pm n}} -\sum_{\pm }
\frac{\alpha(x_n)}{1/x-x_{\pm n}}
 \\
\delta q_2=-\delta q_{9} = + \sum_{\pm } \frac{\alpha(x_n)}{x-x_{\pm
n}}  -\sum_{\pm } \frac{\alpha(x_n)}{1/x-x_{\pm n}}
\eeqa
yielding
\beqa
\delta E= \sum_{n=\pm} \frac{x_n^2+1}{x_n^2-1}=\sum_{\pm n}
\sqrt{1+\frac{n^2}{\omega^2}} \la{w2}
\eeqa
which is again the same result as found in \cite{BMN}.
\subsubsection{Fermionic excitations}
As for $CP^3$ here we also have two types of fluctuations
corresponding to the first and second lines in figure
\ref{excitations}.
We star by considering a representative of the first line. For example
let us focus on a pole from $q_1$ to $q_5$ (and thus automatically
also from $q_6$ to $q_{10}$). We have
\beqa
&&\delta   q_1=-\delta   q_{10}= +\sum_{\pm } \frac{\alpha(x_n)}{x-x_{\pm n}} \\
&&\delta     q_5=-\delta   q_{6}=  - \sum_{\pm }
\frac{\alpha(x_n)}{x-x_{\pm n}}-\sum_{\pm }
\frac{\alpha(x_n)}{1/x-x_{\pm n}}\\
&&\delta   q_2=-\delta   q_{9} = -\sum_{\pm } \frac{\alpha(x_n)}{1/x-x_{\pm n}}
\eeqa
giving
\beqa
\delta E= \sum_{n=\pm} \frac{x_n^2+1}{2(x_n^2-1)}=\sum_{\pm n}
\sqrt{\frac{1}{4}+\frac{n^2}{\omega^2}} \la{w1}
\eeqa
For a fluctuation in the second line, say the last one, we have
\beqa
&&\delta   q_1=-\delta q_{10}= -\sum_{\pm }
\frac{\alpha(x_n)}{x-x_{\pm n}} - \frac{A x}{2g(x^2-1)} \\
&&\delta   q_2=-\delta  q_{9}= +\sum_{\pm }
\frac{\alpha(x_n)}{1/x-x_{\pm n}} - \frac{A x}{2g(x^2-1)} \\
&&\delta     q_3=-\delta  q_{8}=  + \sum_{\pm }
\frac{\alpha(x)}{x-x_{\pm n}}     \\
&&\delta    q_4=-\delta  q_{7} = -\sum_{\pm } \frac{\alpha(1/x)}{1/x-x_{\pm n}}
\eeqa
so that pole synchronization gives
\beq
A=\sum_{\pm}\frac{1}{x_{\pm n}^2-1}
\eeq
and then from the large $x$ asymptotics we read the energy shift
\beqa
\delta E= \sum_{\pm n} \frac{x_{n}^2+3}{2(x_n^2-1)}=\sum_{\pm n}
\sqrt{1+\frac{n^2}{\omega^2}}-\frac{1}{2}
\eeqa
This completes the computation of the spectrum of the superstring
around the BMN classical solution. All frequencies coincide with those
found in \cite{BMN}.
\section{Chern-Simons curve}
In the scaling limit where the Bethe roots scale with the number of
spin chain sites, the two loop Bethe equations in \cite{MZ} can be
recast as \cite{C1,C4,C6}
\beqa
\frac{1}{z}+2\pi n_u&=&2\, \sG_u-G_r \\
\frac{1}{z}+2\pi n_v&=&2\, \sG_v-G_r \\
2\pi n_r&=&2\, \sG_r-G_v-G_u-G_w \\
2\pi n_w&=&G_r-G_s \\
2\pi n_w&=&2\,\sG_w-G_s
\eeqa
In these five equations $z$ belongs to the several disjoint supports
where the Bethe roots $u$, $v$, $r$, $s$, $w$ condense, respectively.
As usual
\beqa
G_u=\sum_{j=1}^{M_u} \frac{1}{Lz-u_j}\,\,\,  , \,\,\,
G_v=\sum_{j=1}^{M_v} \frac{1}{Lz-v_j} \,\,\, ,\,\dots
\eeqa
and the slash means the average of function above and bellow the cut
resulting from the condensation of the Bethe roots. In this limit the
spin chain can be described by a (super symmetric) Landau-Lifshitz
model and the corresponding algebraic curve can be compared with the
curve described in the previous section. Indeed all the $5$ nested
Bethe equations nested can be turned into the statement that the
quasimomenta
\beqa
&&q_1=-q_{10}=\frac{1}{z}-G_w \nn   \\
&&q_2= -q_9\,\,=\frac{1}{z}+G_w-G_s \nn \\
&&q_3=-q_8\,\,=\frac{1}{z}\qquad\,\,\,\,-G_s+G_r  \\
&&q_4=-q_7\,\,=\frac{1}{z}\qquad\,\,\,\,\qquad\,\,\,\,-G_r+G_u+G_v\nn \\
&&q_5=-q_6\,\,=  \qquad\,\,\,\qquad\,\,\,\,\qquad\,\,\,\qquad-\,G_u+G_v  \nn
\eeqa
form a ten-sheeted Riemann surface.
The several properties of these quasimomenta follow trivially from the
definition of the quasimomenta, see \cite{C1,C2,C3,C4,C5,C6}.  This
curve can be depicted as in figure \ref{curve} provided we drop the
unit circle. The energy of the YM solutions is then given by
\beq
E=\sum_{i=1}^{M_u}\frac{\lambda^2}{u_i^2}+\sum_{i=1}^{M_v}\frac{\lambda^2}{v_i^2}
\eeq
%

let us now consider the degeneration of the string previous algebraic
curve in the Frolov-Tseytlin limit.  In the spectral representation
the analytical properties of the quasi-momenta $q_i$ could be
summarized in the following way (see \cite{C3,C4,C6} for similar
decompostions)
to the function
strong coupling -- see \cite{BMN}.
Finally let us mention that while in the $AdS_5/CFT_4$ we had a three
loop mismatch here we the missmatch is obviously at leading order due
to the function
$f(\lambda)$ interpolating the dispersion relation between weak and
strong coupling -- see \cite{BMN}. In the former case it was
understood to be perfectly expectable and a generic feature of non
commutative character the order of limits involved in the
computations. Thus, here there is nothing worrisome about the mismatch
being already at leading order. On the contrary, this is what we
should expect in a generic situation.

\section*{Acknowledgments}
We would like to thank J.~Minahan and K.~Zarembo for many useful discussions, for introducing us to the subject of integrability in the $AdS_4/CFT_3$ duality and for collaboration in earlier stages of this project. We are also grateful to V.~Kazakov, I.~Kostov, D.~Serban and I.~Shenderovich for useful comments. PV is funded by the Funda\c{c}\~ao para a Ci\^encia e Tecnologia fellowship {SFRH/BD/17959/2004/0WA9}. NG was partially supported by RSGSS-1124.2003.2, by RFFI project grant 06-02-16786 and ANR grant INT-AdS/CFT (contract ANR36ADSCSTZ). A part of this work was done during NG's stay at Les Houches Summer School, which he thanks for  hospitality




\end{document}